\documentclass[12pt,preprint,flushrt]{aastex}

\shorttitle{FIELD CONDITION}
\shortauthors{KOYAMA \& INUTSUKA}

\begin{document}            

\title{The Field Condition: A New Constraint on Spatial Resolution in
Simulations of 
the nonlinear Development of Thermal Instability}

\author{Hiroshi Koyama\altaffilmark{1,2}
 and
Shu-ichiro Inutsuka\altaffilmark{3}}

\altaffiltext{1}{Astronomical Data Analysis Center, 
National Astronomical Observatory, Mitaka, Tokyo 181-8588, Japan}

\altaffiltext{2}{Current address: Department of Earth and Planetary Sciences, 
Kobe University, Kobe 657-8501, Japan; 
hkoyama@kobe-u.ac.jp}

\altaffiltext{3}{
Department of Physics, Kyoto University, Kyoto 606-8502, Japan;
inutsuka@tap.scphys.kyoto-u.ac.jp}

\begin{abstract}
We present the dynamics of a thermally bistable
 medium using one-dimensional numerical calculations, including cooling,
 heating, 
 thermal conduction, and physical viscosity. 
We set up a two-phase
 medium from a thermally unstable one-phase medium and follow its
 long-term evolution.
To clarify the role of thermal conduction, 
 we compare the results of the two models, 
 with and without thermal conduction.
The calculations show that 
 the thermal conduction helps to generate the kinetic
 energy of translational motions of the clouds.
Next, we focus on spatial resolution because we have to resolve the Field 
 length $\lambda_{\rm F}$, which is the
 characteristic length scale of the thermal conduction.
The results show convergence only when thermal
 conduction is included and a large enough cell number is used. 
We find it necessary to maintain a cell size of less
 than $\lambda_{\rm F}/3$ to achieve a converged motion in the two-phase
 medium. 
We refer to the constraint that $\lambda_{\rm F}/3$ be resolved
 as the ``Field condition''. 
The inclusion of thermal conduction to satisfy the Field condition is 
 crucial to numerical studies of thermal instability (TI) and may be
 important for studies of 
 the turbulent interstellar two-phase medium:  
 the calculations of TI without thermal conduction 
 may be susceptible to contamination by artificial phenomena that do not 
 converge with increasing resolutions.
\end{abstract}


\section{Introduction}

The physical properties of turbulence in an interstellar medium (ISM) are 
 considered to be important to the theory of star formation 
 (Nakano 1998; Mac Low \& Klessen 2003) 
and the Galactic ISM (Elmegreen 1999).
Arons \& Max (1975) proposed that molecular cloud turbulence may be
primarily in Alfve\'nic motions, because for linear-amplitude waves,
no compressions are involved.
However, the results of numerical magnetohydrodynamic simulations 
 showed that
 the dissipation time of turbulence is of the order of 
 the flow-crossing time or less (Stone, Ostriker, \& Gammie 1998).
On the other hand, 
the triggering of thermal instability (TI) in ISM 
by strong compressions,
and the subsequent production of small-scale structures, has also been studied
 (Parravano 1987; 
  Hennebelle \& P\'erault 1999, 2000; Koyama \& Inutsuka 2000,
  hereafter KI00). 
Recently, Koyama \& Inutsuka (2002, hereafter KI02) proposed 
 a mechanism based on TI 
 to generate and maintain tiny clouds with supersonic velocity
 dispersion in a shock-compressed layer of ISM 
 by performing numerical calculations with radiative heating, cooling, 
 and thermal conduction.
Kritsuk \& Norman (2002) studied TI-induced turbulence in ISM
 numerically 
 and measured the monotonic decay of turbulence. 
In contrast,
 V\'azquez-Semadeni, Gazol, \& Scalo (2000)  have done large-scale
 numerical simulations of TI, with and without prescribed energy
 injection (forcing), and conclude that 
 TI is of minor importance to the resulting spatial and probability
 distributions of the density field if the interstellar turbulence is
 driven by stellar and other energetic sources.

Thermal conduction can stabilize TI when the perturbation wavelength is 
 smaller than the Field length 
 $\lambda_{\rm F}=(KT/\rho^2\Lambda)^{1/2}$ (Field 1965, see
 description below eq. [4]).
If the thermal conduction is not included in the numerical analysis, 
 the perturbations of smaller wavelengths (down to the size of a cell)
 become unstable so that, in principle, the result depends on 
 the size of the cell in a numerical model. 
The Field length also determines the width of the intermediate region 
 between the cold medium and surrounding warm medium, 
 in which heating and cooling by thermal conduction balance 
 the radiative heating and cooling (Begelman \& McKee 1990).
This intermediate region is lost in calculations without 
 thermal conduction. 
Thus, the inclusion of thermal conduction is necessary 
 to describe the physical evolution of TI and to resolve 
 the critical length scale of TI.

Similar circumstances can be found in hydrodynamical 
 evolution with self-gravity, where the Jeans length plays 
 a critical role.
Truelove et al. (1997, 1998) studied  
 the convergence of numerical solutions in 
 isothermal gravitational collapse problems 
 and found it necessary to keep the ratio of cell size to a Jeans length
 $(\Delta x/\lambda_{\rm J})$ below 0.25 to avoid artificial 
 fragmentation. 
They termed this constraint on Jeans instability 
 the ``Jeans condition''. 
Bate \& Burkert (1997) found a similar criterion for particle methods. 

In this Letter, we investigate 
 the convergence property of numerical solutions in 
 TI-induced two-phase gas dynamics. 
We show how thermal conduction plays a role 
 in the nonlinear development of the TI, 
 and find a necessary condition for the numerical analysis 
 of the nonlinear development of the TI. 

\section{Simulations}

We solve the following hydrodynamic equations:

\begin{equation}
\frac{\partial \rho}{\partial t}+\frac{\partial}{\partial x}(\rho v)=0,
\end{equation}
\begin{equation}
\frac{\partial}{\partial t}\rho v+\frac{\partial}{\partial x}
\left(P+\rho v^2-\frac{4\mu}{3}\frac{\partial v}{\partial x}\right)=0, 
\end{equation}
\begin{equation}
P=\frac{\rho k_{\rm B}T}{m},
\end{equation}
\begin{eqnarray}
\frac{\partial E}{\partial t}+\frac{\partial}{\partial x}\left(
(E+P)v
-\frac{4\mu}{3}v\frac{\partial v}{\partial x}
-K\frac{\partial T}{\partial x}
\right)= 
\frac{\rho}{m}\Gamma-\left(\frac{\rho}{m}\right)^2\Lambda(T).
\end{eqnarray}
In the above equations,
$\rho$ is the mass density of the gas,
$v$ is the velocity of the fluid elements, $P$ is the gas pressure,
$m$ is the average particle mass,
$k_{\rm B}$ is the Boltzmann constant,
$E=P/(\gamma-1)+\rho v^2/2$ is the total energy per volume,
and $K$ is the coefficient of thermal conductivity.
For the range of temperatures considered,
the dominant contribution to thermal conductivity is that
of neutral atoms, for which
$K = 2.5 \times 10^3 T^{1/2}\,{\rm cm^{-1} K^{-1} s^{-1}}$
(Parker 1953).
We assume that the gas consists of monoatomic molecules, and
 the ratio of the specific heats is $\gamma=5/3$.
The relation between kinetic viscosity $\mu$ and thermal conductivity $K$
 is characterized by a Prandtl number:
\begin{equation}
p=\frac{\gamma}{\gamma-1}\frac{k_{\rm B}}{m}\frac{\mu}{K}.
\end{equation}
In this Letter, we assume the Prandtl number $p=2/3$ for monoatomic
 molecules.
For the heating/cooling function, we adopt a simple fitting formula 
(see eqs. [4] and [5] in KI02)  
based on detailed thermal and chemical calculations (KI00).
 
The adopted
hydrodynamical method is the one-dimensional Eulerian
code based on the second-order Godunov
scheme (van Leer 1979). We use explicit time integration for
cooling and heating.
We also use explicit time integration for thermal
conduction, and physical viscosity with spatially second-order accuracy.

\subsection{Results}
Figure 1 shows the typical numerical results.
The length of the computational domain is $L=$4.8 pc. 
The boundary conditions are periodic.
In this case, cell number of 16,384 is used.
Upper panel shows the initial density.
The initial condition is a static unstable equilibrium
 ($\rho/m=2.93$cm$^{-3}$, $T=756$ K) 
 with a superposition of isobaric density perturbations with wavelengths
 in the range 0.15 pc $\le |\lambda| \le$ 4.8 pc.
Here, we introduce the characteristic length scale of TI, 
 $\lambda_{\rm a}$ defined as the sound speed times the cooling
 time. 
We can say that perturbations with $\lambda \ge \lambda_{\rm a}$
 correspond to the long-wavelength mode; those with
 $\lambda_{\rm F} \le \lambda \le \lambda_{\rm a}$, to the
 intermediate-wavelength mode.
It has been established that the intermediate-wavelength perturbations
 evolve almost 
 quasistationarily, maintaining quasi-isobaricity throughout their
 evolution.
On the other hand, the long-wavelength perturbations cool before they
 can equate pressures and therefore develop larger pressure gradients,
 which then cause them to evolve highly dynamically 
 (Meerson 1996; Burkert \& Lin 2000; S\'anchez-Salcedo et al. 2002).
The $\lambda_{\rm a}$ is 0.5 pc in the initial parameters, and thus 
 our simulations include both modes. 
The middle panel in Figure 1 shows the result of cooling+
viscosity models (hereafter model CV). 
In this case, 17 clouds are formed because smaller wavelength
 perturbations have larger growth rates of TI.
The lower panel in Figure 1 shows the result of cooling+
conduction+viscosity models (hereafter model CCV).
In this case, however, only nine clouds are shown.
Note that all perturbations grow at the initial stage
 because their wavelengths are larger than the Field length 
 $\lambda_{\rm F}$ = 0.008 pc in the initial parameters. 
The decrease in the cloud number is the consequence of coalescence. 
Across the interface between the two-phase medium,
 the conductive heat flux changes the local pressure distribution, 
 and the induced pressure gradient produces gas motion. 
Thus, the coalescence is driven by thermal conduction. 

Figure 2 shows the time evolution of kinetic energy per volume. 
In the early phase ($0 \le t \le 3$ Myr),
 the linear perturbation grows.
The smaller wavelength perturbations have larger growth rates
 and hence evolve rapidly and attain a quasiequilibrium state in
 short timescales.
In the late phase ($t \ge 5$ Myr),
 the kinetic energy in model CCV (solid line) is 1 order of magnitude 
 larger than the case in model CV (dotted line).
This shows that 
 the kinetic energy of the translational motions are enhanced by the
 effect of thermal conduction.
This shows how thermal conduction plays a role in the nonlinear 
 development of TI. 

The conductive-length scale is characterized by the Field length. 
The Field length is a function of temperature and density,
 and hence should be defined locally: 
 it varies spatially with a range of 0.6--0.003 pc in this problem. 
We measured the number of clouds and the maximum local Mach number with
 various resolutions (Table 1). 
In model CCV, both numbers converge when 
 the cell size, $\Delta x$, is smaller than 0.39$\lambda_{\rm F,min}$.
In model CV, on the other hand, 
 the results with higher resolutions have larger numbers of clouds and
 smaller Mach numbers. 
These resolution-dependent results indicate that model CVs have a
numerical problem.
In the absence of thermal conduction, 
 a spatially smooth transition from a cold to a warm phase is
 impossible, and the spatial distribution of temperature tends to produce
 discrete jumps from a cold phase to a warm phase. 
When the cloud moves, 
 the advection between the Eulerian cells produces cells with intermediate and
 thermally unstable temperatures.
Each of these cells becomes unstable, because the most unstable wavelength
 becomes infinitesimal in the absence of thermal conduction.
This thermally unstable cell generates unphysical unstable motion.
This artificial noise cannot be removed by increasing 
 the number of cells. 

The above results suggest that 
 conductive length scale characterized by the local Field length 
 should be resolved in physically acceptable numerical solutions. 
Figure 3 shows the convergence test for density distribution at $t=8$ Myr.
The error function is defined as
\begin{equation}
\epsilon_N = 
\sum_{i}^{N} \Delta x_{N} |\rho_N(x_i,t)-\rho(x_i,t)|,   
\end{equation}
where $N$ is the number of cells and $i$ is the label of cells. 
We use the result with 32,768 cells as a reference solution:
 $\rho(x,t)=\rho_{32,768}(x,t)$.  
The filled circles denote model CCV, and the open boxes denote model CV. 
Because the numerical scheme we use is spatially a second-order scheme, 
 the numerical errors should decrease with the square of cell length.
In the calculations with 256, 512, and 1024 cells, the errors in models CV
 and CCV 
 are comparable, because neither model resolves Field length.
The errors with 4096, 8192, and 16,384 cells in model CCV show second-order
 convergence, while the amount of error in model CV does not decrease 
 toward second-order accurate solutions.
The calculation with 4096 cells uses a uniform cell length, $\Delta
 x_{4096}=0.39\lambda_{\rm F,min}$. 
Therefore, 
 we conclude that 
 we should always use no less than three cells per local Field
 length to 
 obtain a physically acceptable solution in the calculation of TI.
We term this new constraint the `Field condition'.
The inclusion of thermal conduction is a necessary but not sufficient
 condition to satisfy the Field condition.

\section{Discussion}

The thickness of the interface characterized by the Field length
 is typically much smaller than the physical sizes of 
 the interstellar clouds.
Thus, the effect of the thermal conductivity is assumed to be 
 negligible in many numerical studies, with the exception of our own.
Our simulations that satisfy the Field condition show 
 more dynamical motions. 
Let us consider the physical meaning of the difference.
For convenience of explanation,
 we assume the steady state of the two-phase medium.
By integrating hydrodynamic equations over the interface, 
 we obtain 
\begin{eqnarray}
\left[\rho v\right]^{\rm cold}_{\rm warm}&=&0, \\
\left[P+\rho v^2\right]^{\rm cold}_{\rm warm}&=&0, \\
\left[(E+P)v\right]^{\rm cold}_{\rm warm}&=&q, 
\end{eqnarray} 
where
\begin{equation}
  q=\int^{\rm cold}_{\rm warm} 
  \left[
  \left(\frac{\rho(x)}{m}\right)\Gamma
  -\left(\frac{\rho(x)}{m}\right)^2\Lambda(T(x))~
  \right]
  dx. 
\end{equation}
Note that the gradients of velocity and temperature are negligible 
 in the region far from the interface
 and therefore
 the net viscosity and thermal conductivity
 vanish in this integral form. 
If the integral $q$ vanishes, the set of the equations is 
 equivalent to the Rankine-Hugoniot jump condition.
This condition enable us to 
 estimate the difference between preshock and postshock states
 without solving the detailed structure. 
In the case of two-phase medium, on the other hand, 
 $q$ does not always vanish:
 $q$ vanishes only when the pressure is equal to the saturated
 pressure (Zel'dovich \& Pikel'ner 1969). 
When the thermal conductivity is not included, the corresponding
 two-phase interface becomes zero-thickness and $q$ vanishes.
This is the most essential physical difference between systems 
 with and without thermal conduction.  
Here we emphasize that the finite value of the integral $q$ should 
 not be ignored --- in spite of the small thickness of the interface ---  
 because this $q$ causes the time dependence in pressure and hence 
 maintain dynamical motion.  
It seems impossible to estimate this integral $q$ 
 without knowing the specific structure between the interface. 
At present, only one method is available for handling this problem --- 
 the use of a high spatial resolution to describe
 the detailed interface structure. 
The need for the spatial resolution of the Field length does not 
 require the spatial resolution of shock fronts 
 because the former is much larger than the latter: 
\begin{equation}
  \lambda_{\rm F}=\sqrt{ \frac{KT}{\rho^2\Lambda} } 
  ~ \sim ~~ l ~ \sqrt{ \frac{t_{\rm c}}{t_{\rm mfp}} } 
   ~~ > ~~ l ~, 
\end{equation}
 where $l$ is the mean free path, 
 and $t_{\rm c}$ and $t_{\rm mfp}$ are 
 the cooling time and mean flight time, respectively.  

\section{Conclusion}
In this Letter, we investigate the dynamical evolution of TI
 using one-dimensional hydrodynamical calculation, 
 including heating, cooling, 
 thermal conduction, and physical viscosity.
We present the results of the convergence tests, 
 with and without thermal conduction.
Our findings are as follows:

1. 
The simulations without thermal conduction show that
 the results are resolution-dependent.
The majority of the clouds are formed in higher resolution runs. 
The maximum local Mach number decrease with increasing resolution.
 
2.
The translational motions survive in the calculations
 with thermal conduction and sufficient spatial resolution.
This shows that thermal conduction helps to generate the kinetic
 energy of the translational motions of the two-phase medium.
We found the maximum local Mach number to be 0.13 at t = 8 Myr
 in the numerical model.

3.
At least three cells are required to resolve the crucial length 
 scale, the Field length, to attain the convergence of 
 the numerical calculations. 
We refer to this constraint as the ``Field condition''.  
The inclusion of thermal conduction is imperative to satisfy
 the Field condition.

\acknowledgements

Numerical computations were carried out on VPP5000 at the Astronomical
Data Analysis Center of the National Astronomical Observatory, Japan.

\begin{table}
\begin{center}
\caption{Results from Various Resolutions at t = 8Myr \label{tbl-1}}
\begin{tabular}{rlcrcr}
\tableline\tableline
\multicolumn{2}{c}{} & \multicolumn{2}{c}{Model CV} 
& \multicolumn{2}{c}{Model CCV} \\
Number of Cells & $\Delta x/\lambda_{\rm F,min}$
& \# of clouds  & $M$\tablenotemark{a} 
& \# of clouds  & $M$\tablenotemark{a} \\
\tableline 
 32768 & 0.0488 & 16 & 0.054 & 9 & 0.134 \\
 16384 & 0.0976 & 16 & 0.075 & 9 & 0.134 \\
  8192 & 0.1953 & 17 & 0.117 & 9 & 0.136 \\
  4096 & 0.3906 & 14 & 0.138 & 9 & 0.167 \\
  2048 & 0.7813 & 11 & 0.168 & 8 & 0.188 \\
  1024 & 1.5625 &  8 & 0.269 & 8 & 0.198 \\
   512 & 3.125  &  6 & 0.212 & 6 & 0.234 \\
   256 & 6.25   &  5 & 0.097 & 6 & 0.176 \\
\tableline
\end{tabular}
\tablenotetext{a}{Maximum local Mach number.}
\end{center}
\end{table}

\clearpage

\begin{figure}
\figurenum{1}
\epsscale{0.8}
 \plotone{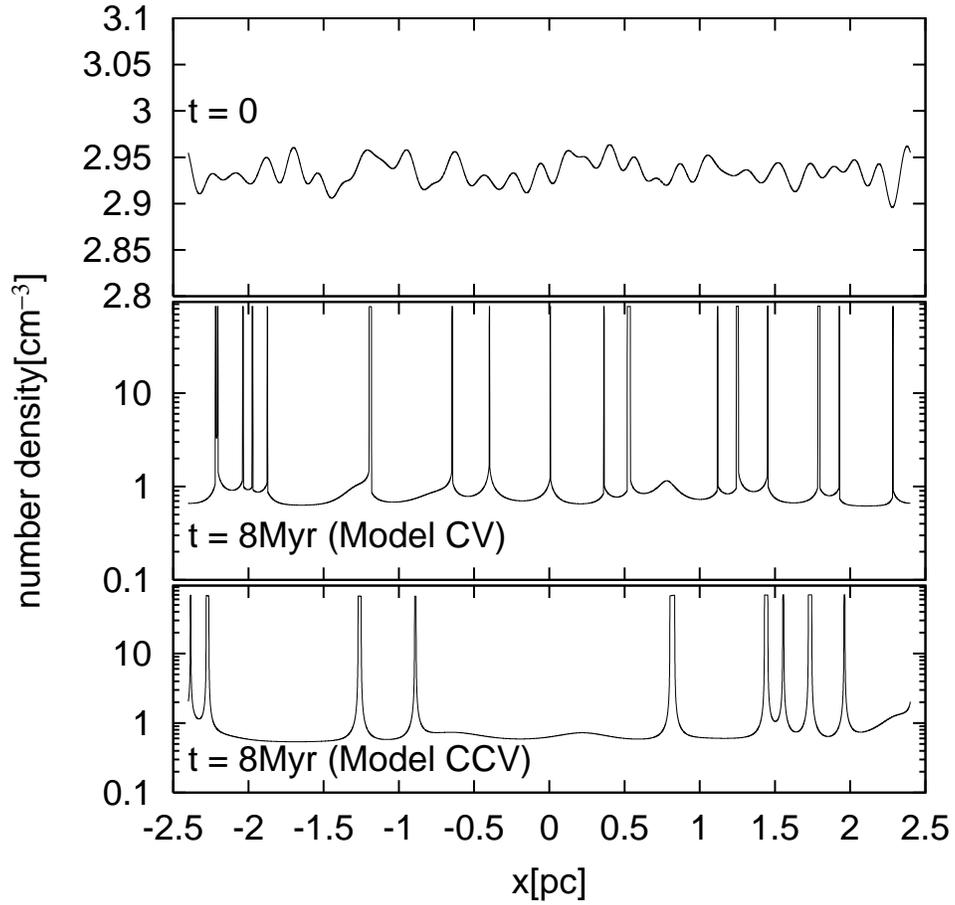}
\caption{Evolution of number density. The number of cells is 16,384.
{\it Top}: initial density profile.
{\it Middle}: Final snapshot for the run without thermal
 conduction (model CV). 
{\it Bottom}: Final snapshot for the run with thermal conduction
 (model CCV). 
}
\label{fig1}
\end{figure}

\begin{figure}
\figurenum{2}
\epsscale{0.95}
 \plotone{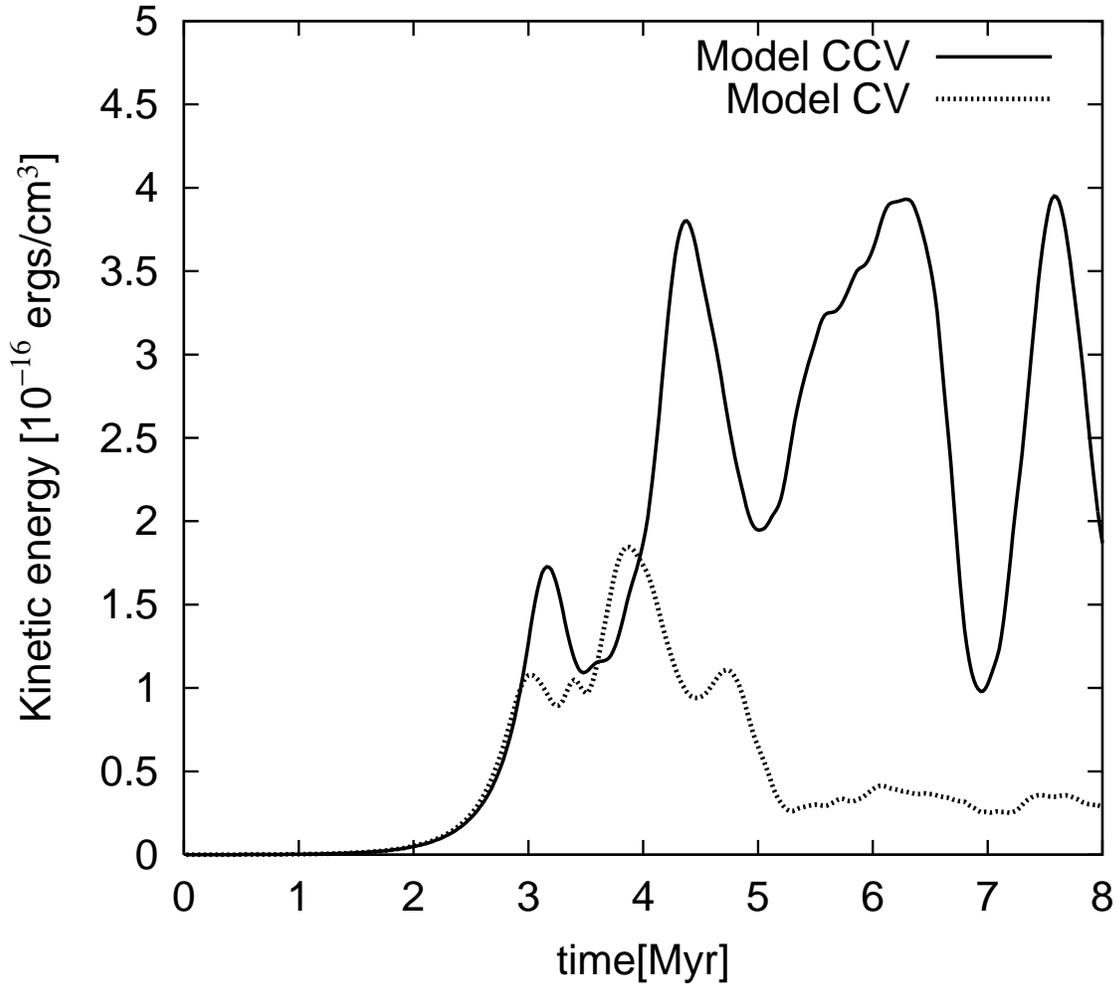}

\caption{
Time evolution of the kinetic energy in models CV (dotted line) and CCV
 (solid line). The number of cells is 16,384.
}
\label{fig2}
\end{figure}

\begin{figure}
\figurenum{3}
\epsscale{0.8}
 \plotone{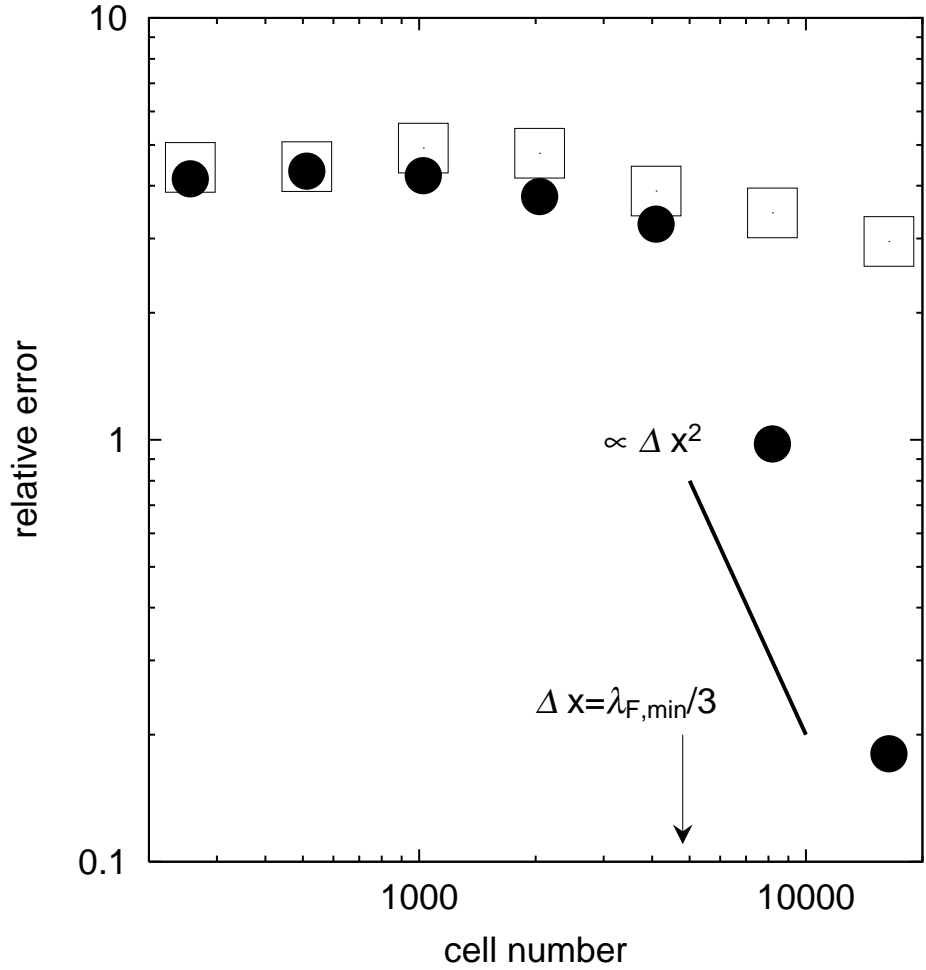}

\caption{
Convergence test for density distribution at $t=8$ Myr. 
The error function is defined by equation (6). 
Model CV ({\it open squares}) and model CCV ({\it filled circles}) are presented. 
}
\label{fig3}
\end{figure}

\end{document}